\documentclass[aps,prl,reprint,superscriptaddress]{revtex4-2}

\usepackage[a4paper, total={174mm,259mm},left=18mm,top=18mm]{geometry}
\usepackage{amsmath,amssymb,amsfonts}
\usepackage{algorithmic}
\usepackage{graphicx}
\usepackage{wrapfig}
\usepackage{caption}
\usepackage{subcaption}
\usepackage{float}
\usepackage{hyperref}
\hypersetup{colorlinks=true,linkcolor=teal,urlcolor=blue,filecolor=red,citecolor=olive}
\usepackage{mathrsfs}
\usepackage{tcolorbox}
\usepackage{textcomp}

\begin{document}
\title{Electron Thermometry}

\author{Joost van der Heijden}
\affiliation{QDevil, Quantum Machines, 2750 Ballerup, Denmark}

\begin{abstract}
    The performance and accuracy of quantum electronics is substantially degraded when the temperature of the electrons in the devices is too high. The electron temperature can be reduced with appropriate thermal anchoring and by filtering both the low frequency and radio frequency noise.
    Ultimately, for high performance filters the electron temperature can approach the phonon temperature (as measured by resistive thermometers) in a dilution refrigerator.
    In this application note, the method for measuring the electron temperature in a typical quantum electronics device using Coulomb blockade thermometry is described.
    This technique is applied to find the readily achievable electron temperature in the device when using the QFilter provided by QDevil.
    With our thermometry measurements, using a single GaAs/AlGaAs quantum dot in an optimized experimental setup, we determined an electron temperature of 28 $\pm$ 2 mK for a dilution refrigerator base temperature of 18 mK.
\end{abstract}

\maketitle
Quantum electronics experiments play a central role in the development of new quantum technologies.
Quantum effects occurring at ultra low temperatures have shown tremendous potential.
Investigating these physical phenomena typically requires the use of dilution refrigerators which allow to cool experimental setups down to phonon temperatures of the order of 10 mK.
The performance of the quantum electronics is, however, also dictated by the temperature of the electrons in the device.
This is typically found to be higher than the phonon temperature normally measured with resistive thermometers such as ruthenium-oxide devices.
To determine the electron temperature, different types of devices and measurement techniques are available \cite{Giazotto06}, such as tunnel junctions between a normal metal and a superconductor (NIS junction) \cite{Rowell76}, superconducting tunnel junctions (SIS junction) \cite{Booth96}, shot noise thermometry \cite{Spietz03}, and thin-film thermometry near the metal-insulator transition \cite{Delinger94}.
In this application note we give a detailed description of the Coulomb blockade thermometry method \cite{Pekola94,Farhangfar97,Mueller13,Maradan14,Meschke16}, as pioneered by J.P. Pekola \cite{Pekola94,Farhangfar97} and based on well established theory \cite{Averin91,Beenakker91}.
For this purpose we use a multi-gate GaAs/AlGaAs single quantum dot.

\section{Method and Instrumentation}\label{sec:method}

\subsection*{Device fabrication}\label{subsec:device}
\begin{figure*}[!ht]
    \centering
    \includegraphics[width=1.00\linewidth]{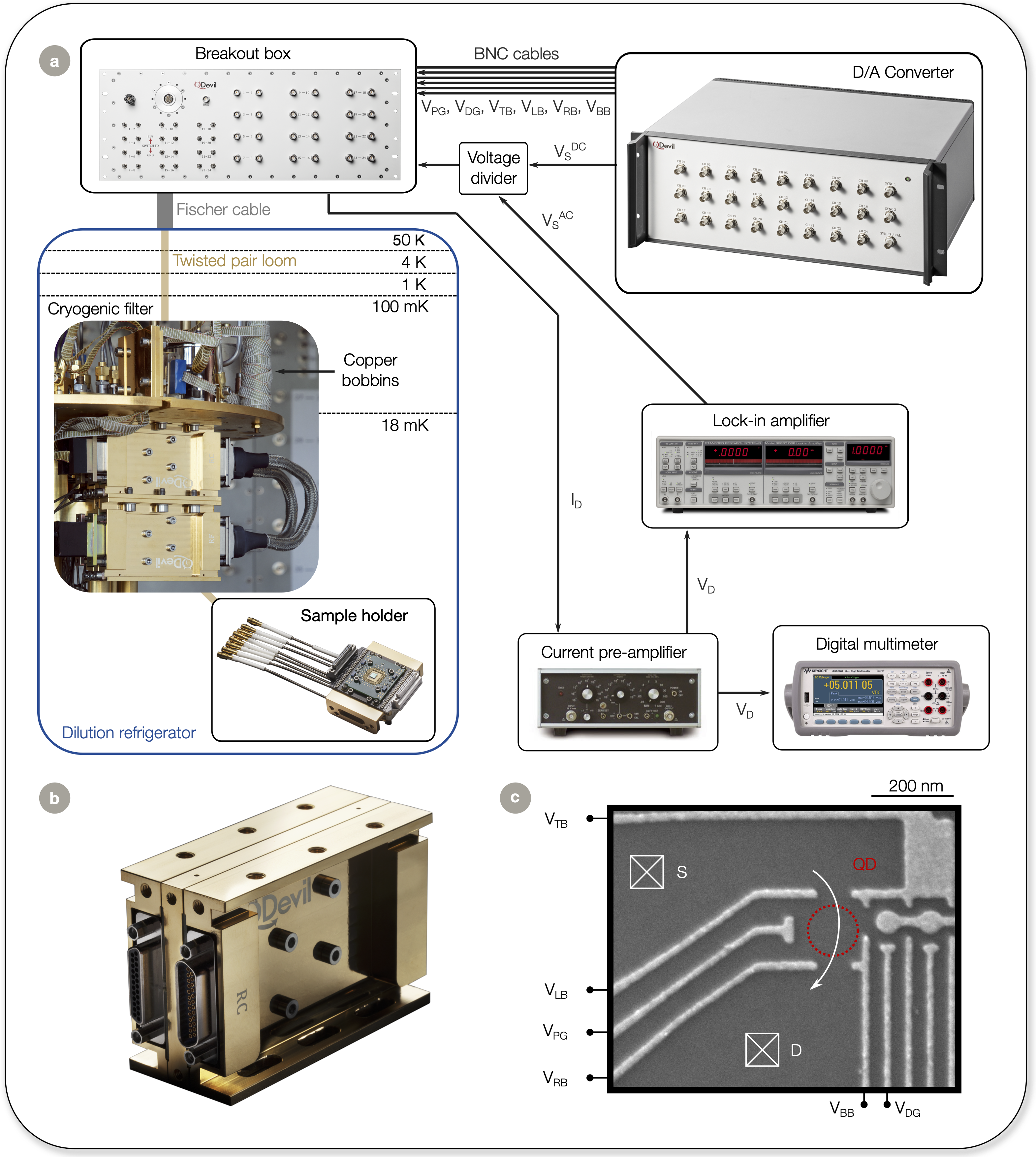}
    \captionof{figure}{\textit{Overview of the experiment. a) Experimental setup, showing the use of a QDevil QDAC D/A converter, a Stanford Research Systems SR830 lock-in amplifier, a QDevil QBox breakout box, an Ithaco 1211 Current preamplifier, a Keysight 34465A digital multimeter, a QDevil QFilter and a QDevil QBoard sample holder. Furthermore the BNC cables, Fischer cable and cryogenic cabling are shown schematically. b) Image of the QDevil QFilter-II, as used in this application note. c) Scanning electron microscope image of the GaAs/AlGaAs device used in this application note. The quantum dot (QD), source (S), drain (D) and plunger gate (PG) are indicated, as well as the barrier gates at the top (TB), bottom (BB) and on the left and right of the quantum dot (RB, LB), and a distant gate (DG). Image courtesy F. Fedele (from F. Fedele thesis \cite{Fedele19}).}}
    \label{fig:1}
\end{figure*}

For the Coulomb blockade thermometry, we make use of a semiconductor single quantum dot device, through which the transport signatures of single electron tunneling can be measured. 
This requires at least a source (S) and a drain (D) electrode and a plunger gate (PG) to control the potential of the quantum dot.
Many different material systems and gate electrode layouts can be used for this purpose \cite{Hanson07}, including commercially available transistors.
In this application note, we make use of a multi-gate quantum dot device defined in a two dimensional electron gas (2DEG), formed in a GaAs/AlGaAs heterostructure.
Carefully shaped metallic gates are used to apply an electrostatic potential and confine the quantum dots laterally.
Although the full device consists of many gates, for this measurement we will only use the gates controlling a single quantum dot, shown in Fig. \ref{fig:1}c.
Next to the source, drain, and plunger gate electrodes, this also includes several barrier gates (TB, BB, LB, RB), which give more control over the electrostatics of the quantum dot.
More details about the fabrication of this device can be found in \cite{Fedele19,Fedele21}.

\subsection*{Experimental setup}\label{subsec:setup}
\begin{figure}[!b]
    \centering
    \includegraphics[width=1.00\linewidth]{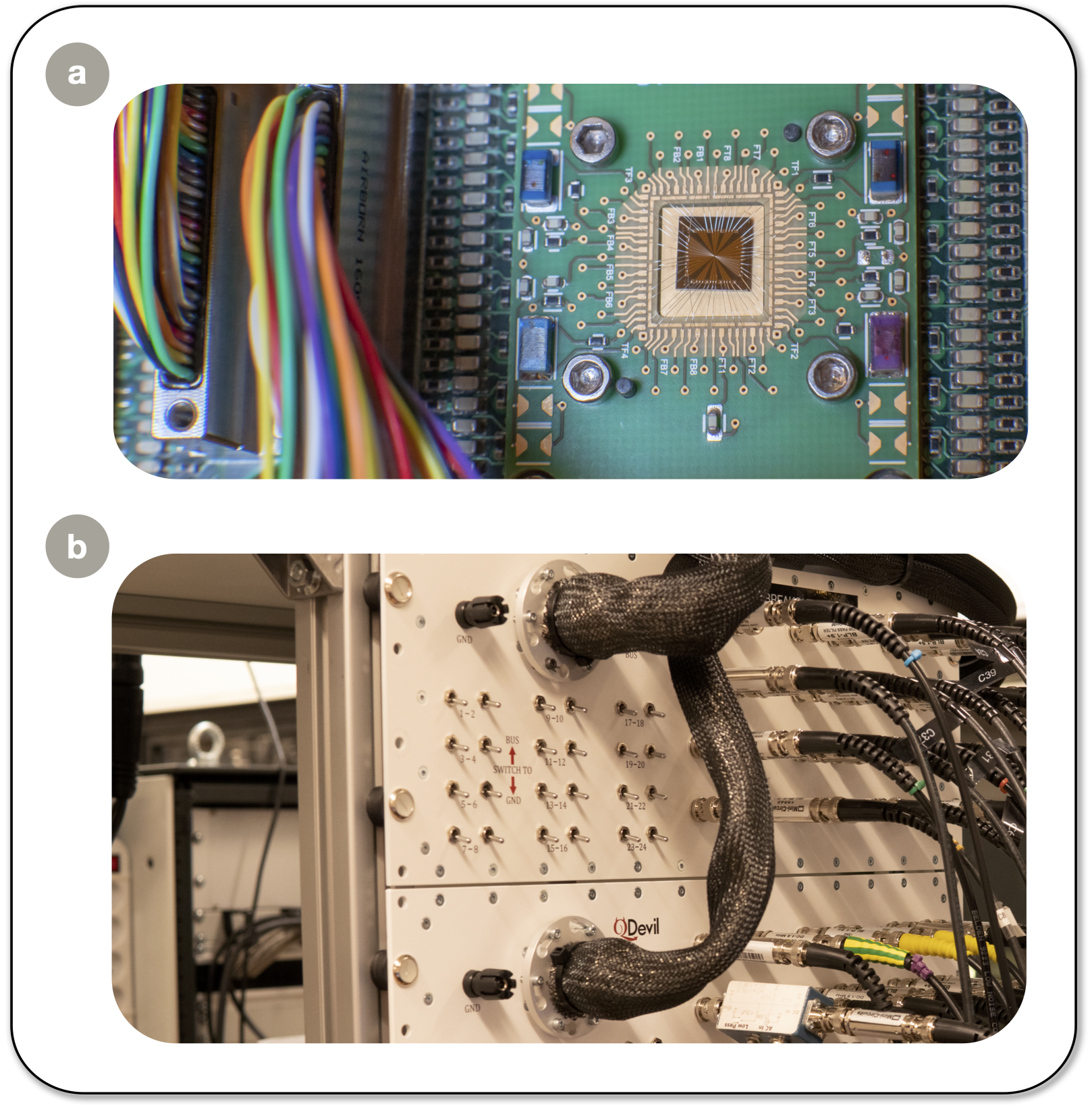}
    \caption{\textit{Photographs of the experiment. a) GaAs sample mounted and wire-bonded in the QBoard, placed in an Oxford puck loading system. b) Grounding of the breakout box through the metal collar system on the Fischer cable connector.}}
    \label{fig:2}
\end{figure}

\begin{figure}[!b]
    \centering
    \includegraphics[width=1.00\linewidth]{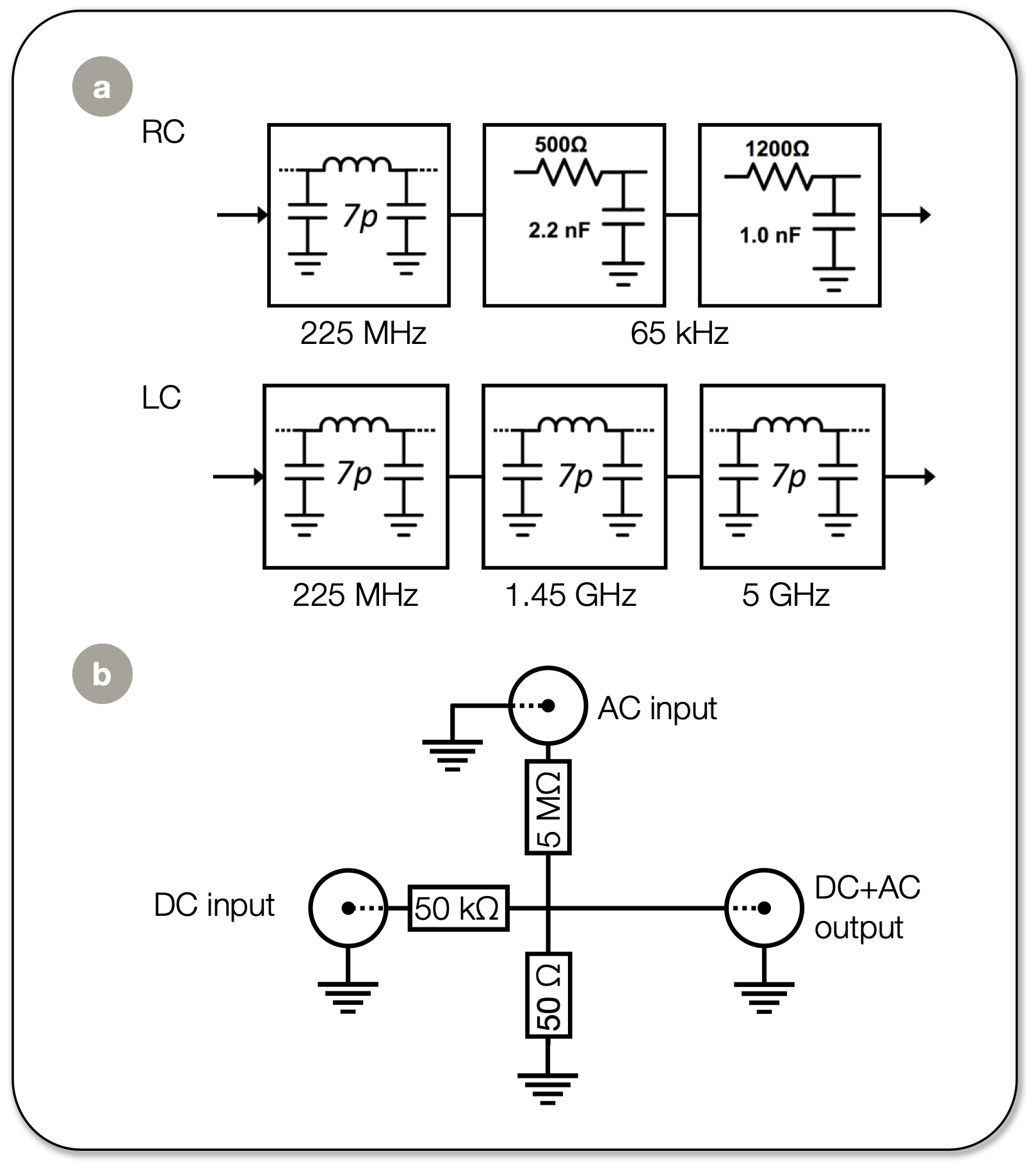}
    \caption{\textit{Electronics schematics. a) Schematic overview of the QFilter RC and LC filter banks. b) Schematic of the voltage divider used to apply DC and AC voltages to the source electrode.}}
    \label{fig:3}
\end{figure}

The quantum dot sample is placed on a sample holder (QBoard by QDevil), which accommodates the connection of the on-chip electrodes to the transmission lines in the cryostat, and ensures a good thermalization of the sample.
For our experiment, we fixed the GaAs chip on a QBoard daughterboard and wire-bonded all gates and ohmic contacts to the bonding pads.
This daughterboard was then placed on a QBoard motherboard, inside a loading puck for the Triton Cryofree dilution fridge bottom-loading system (see Fig. \ref{fig:2}a).
After loading, the QBoard connects to two 24-wire twisted-pair looms, used for DC signals, inside the dilution fridge.
This setup is illustrated in Fig. \ref{fig:1}.
A list of the instrumentation is provided in Appendix A.
Hot electrons enter the dilution refrigerator at room temperature and need to be thermalized at the thermal stages at lower temperatures.
To lower the electron temperature for the experiment, special care should be taken on two points when placing the DC wiring in the cryostat.
First, the wiring should be thermally anchored at each temperature stage of the cryostat.
Here we achieve this with the use of copper bobbins. Second, final thermalization below 1K and filtering of high frequency noise is achieved with the use of the QFilter-II, provided by QDevil.
This is a two part filter consisting of a filter bank with RC low pass stages, which filters the electronic noise from 65 kHz up to the GHz regime and thermalizes the electrons, and a filter bank with LC low pass stages, which filters frequencies from 225 MHz to the THz regime (see Fig. \ref{fig:3}a).
Transmission data and further technical details can be found in the QFilter-II product brochure.
Additional RC filtering of the DC lines occurs on the QBoard.
On the outside of the cryostat, it is useful to split the two 24-wire cables into separate connectors for each electrode to simplify the electrostatic control.
In this experiment we make use of two QBox breakout boxes, provided by QDevil.
They are connected to the cryostat with double shielded low-noise 24 channel twisted pair cables, with Fischer connectors on each end.
Each channel is connected to a BNC connector on the breakout box.
To prevent any ground loops (see the section test of background noise), only a single ground should be used in the experiment.
We choose to use the ground of the magnet power supply of the cryostat as common ground.
To ensure the propagation of this common ground, the extended shielding of the 24 channel cable is connected to the conductive flange of the QBox and the cryostat with metal collars (see Fig. \ref{fig:2}b), while these breakout boxes are otherwise isolated from any other ground carrying equipment or equipment racks.
To apply computer-controlled voltages to the sample electrodes, we choose to use the 24-channel QDAC, provided by QDevil, featuring 20-bit resolution and output noise under 20 nV/Hz$^{1/2}$, which we connect with BNC cables to the breakout box.
For extra filtering, we insert additional 1.9 MHz BNC low pass filters (BNP1.9+ from Mini-Circuits) for all applied voltages on the breakout box.
Again, to prevent any ground loops, the QDAC is only grounded through the BNC cables, powered by an external isolated power supply, and otherwise isolated from any other ground carrying equipment or equipment racks.
To measure the single electron tunneling current in the quantum dot, the channel connected to the source or drain of the device should be connected to a current pre-amplifier and can then be measured in a variety of ways, for instance using a digital multimeter.
In our experiment we use an Ithaco 1211 preamplifier with a conversion of 10$^7$ V/A, and a time constant of 30 ms.
We measure the resulting voltage with a Keysight 34465A 6\textonehalf{} digit multimeter.

Furthermore, to measure the differential conductance in the quantum dot, a lock-in amplifier can be used to apply a small AC voltage on the source or drain electrode.
The resulting oscillation in current will then allow for a sensitive measurement of the differential conductance with the lock-in amplifier. 
Here we use a Stanford Research Systems SR830 lock-in amplifier with a 224 Hz, 3 $\mu$V peak-peak AC excitation voltage (created with an output of 300 mV and the use of a 1/10$^5$ voltage divider, as shown in Fig. \ref{fig:3}b).
We use the direct output of the Ithaco preamplifier (bypassing its low pass filter) as the signal input of the lock-in amplifier to sense the AC current signal.
We use a sensitivity of 500 $\mu$V, a time constant of 300 ms and a filter roll-off of 24 dB.
We convert the measured signal to a differential conductance by:
\begin{equation}\label{Eq1}
    g_{diff} = \frac{dI}{dV_{bias}} = S_{LI}\frac{R_Q}{2G_{pa}V_{ex}},
\end{equation}
Where g$_{diff}$ is the differential conductance in units of e$^2$/h, S$_{LI}$ is the signal from the lock-in amplifier in V, R$_Q$ is the quantum of resistance h/2e$^2$ (around 12 k$\Omega$), G$_{pa}$ is the gain of the Ithaco preamplifier in V/A and V$_{ex}$ is the excitation voltage in V.
All experiments shown in this application note are taken at zero magnetic field.
Furthermore, to execute the experiments, we use the python-based data-acquisition framework of QCoDeS \cite{Nielsen24}.

\subsection*{Test of background noise}\label{sec:noise}
Prior to the experiments, as a good practice, we investigated the noise background with a current measurement in deep Coulomb blockade (as outlined in \cite{Churchill12}).
For this purpose, we set the Ithaco gain to 10$^8$ V/A and measured its direct output with a spectrum analyzer.
Here we found the 50 Hz noise peak to be 180 $\mu$V$_{rms}$, lower than the 300 $\mu$V$_{rms}$ limit for low electron temperature measurements as mentioned in \cite{Churchill12}.
Furthermore, by sequentially switching equipment on and off, we found that the 50 Hz noise peak could be further reduced to 80 $\mu$V$_{rms}$ by disconnecting the dilution fridge resistive thermometry readout circuit and down to 25 $\mu$V$_{rms}$ when the Ithaco pre-amplifier was switched from line power to battery power.

\begin{figure}[!b]
    \centering
    \includegraphics[width=1.00\linewidth]{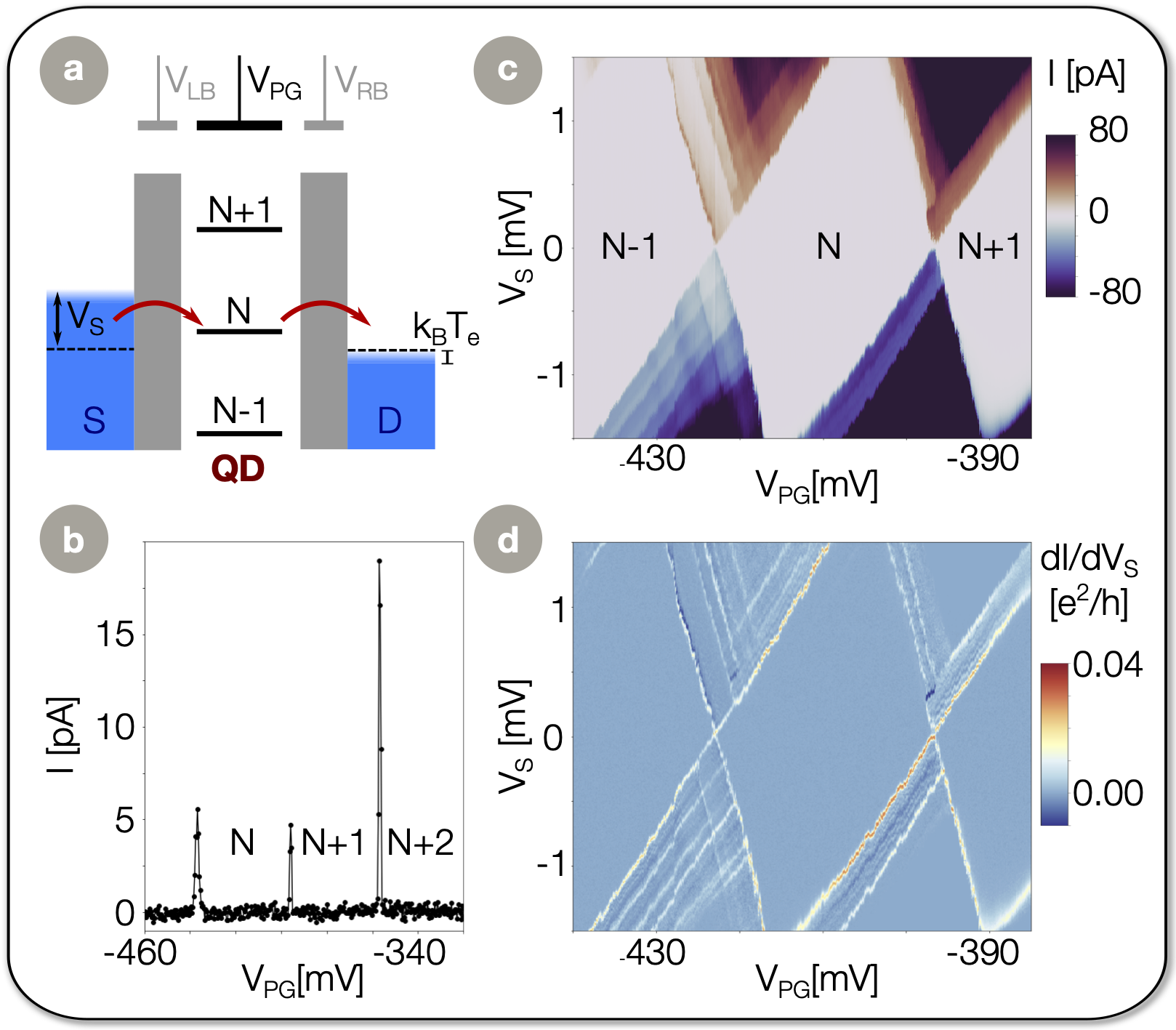}
    \caption{\textit{Coulomb blockade measurements. a) Schematic overview of the Coulomb blockade thermometry measurement. b) Plunger gate voltage trace at a 100 $\mu$V source voltage, showing peaks in current where the single electron tunneling occurs. c) Current measured through the quantum dot as a function of the voltages applied to the plunger gate and source contacts. b) Differential conductance map measured simultaneously with c).}}
    \label{fig:4}
\end{figure}

\subsection*{Measurement of quantum dot signatures}\label{subsec:qd}
In order to operate the specific device in the experiment as a single quantum dot electron thermometer, schematically shown in Fig. \ref{fig:4}a, we first apply negative voltages of -1.2 V (TB) and -0.6 V (BB) to the gates forming the outer barrier of the quantum dot as shown in Fig. \ref{fig:1}c.
This creates a potential barrier and thus isolates the single quantum dot from the rest of the device.
Next, a small bias voltage is applied to the source contact, in order to create a current between the source and drain contact which is grounded through the Ithaco current pre-amplifier.
Then the barrier gates of the quantum dot are tuned to the point where they almost pinch off the current, about 10-50 mV from the pinch-off voltage.
This is the regime where single electron transport occurs, which is best visualized in a scan of the plunger gate voltage (see Fig. \ref{fig:4}b).
The peaks of current indicate single electron tunneling through the quantum dot, while in between these peaks regions of Coulomb blockade are observed.
By extending this measurement to a 2D scan of both the source and plunger gate voltage, the Coulomb blockade region forms a diamond shape, as shown in Fig. \ref{fig:4}c.
This is commonly referred to as a Coulomb diamond and is a typical signature of single electron tunneling in a quantum dot.
Using the lock-in amplifier we measure the differential conductance at the same time, as can be seen in Fig. \ref{fig:4}d.
These measurements are taken as line scans of the source voltage, while stepping the plunger gate voltage.
We use these measurements to extract information about the quantum dot, such as the plunger gate lever arm and the relation between the relevant energy scales, which we require to calibrate the electron temperature measurements later.

\begin{figure}[!b]
    \centering
    \includegraphics[width=1.00\linewidth]{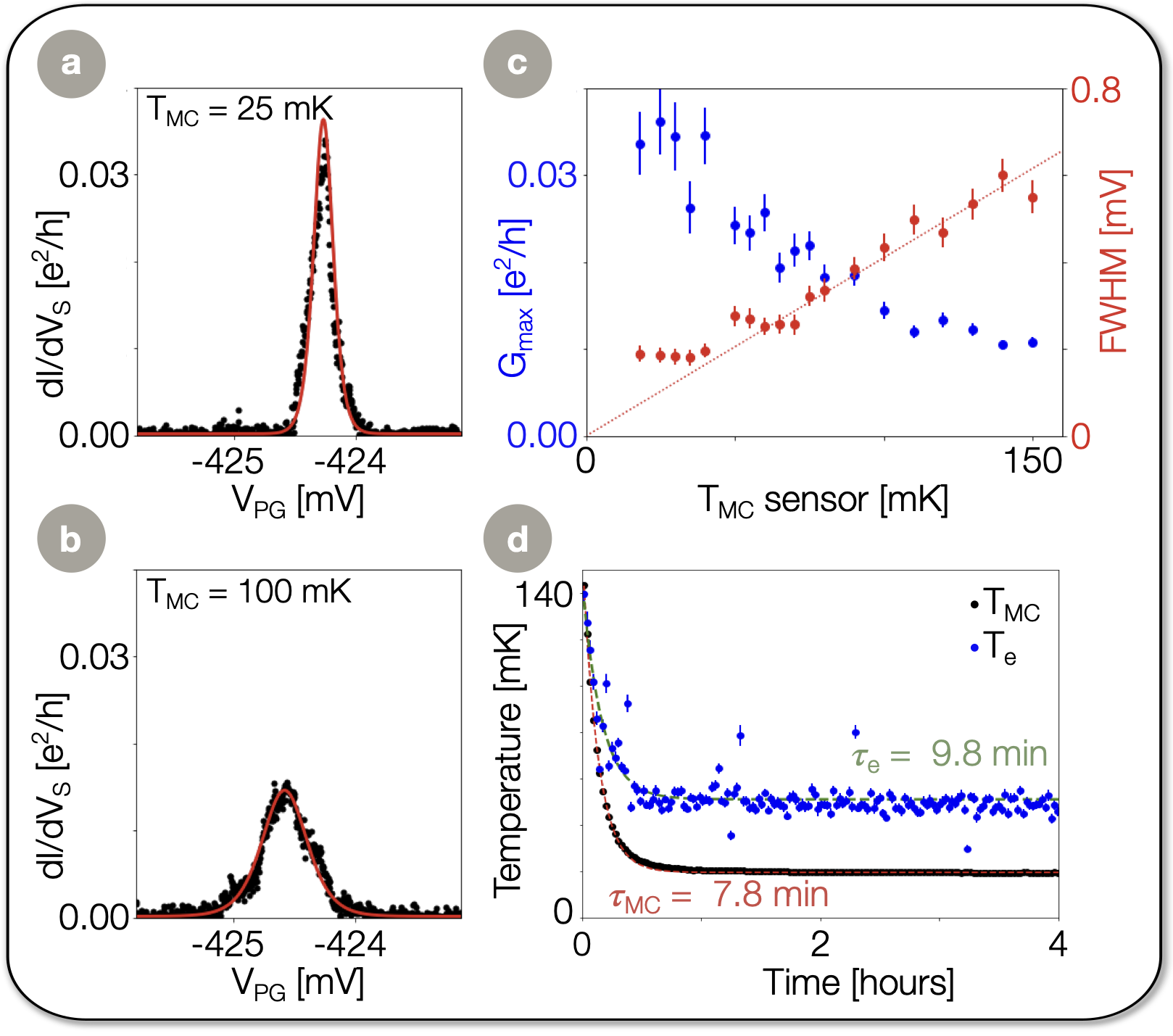}
    \caption{\textit{Temperature dependence of the zero bias conductance peak. a) and b) Data and fit with Eq. \ref{Eq3} of the zero-bias conductance peak, at a mixing chamber temperature sensor measurement of a) 25 mK and b) 100 mK. c) Amplitude and full width half maximum (FWHM) of these as a function of the measured mixing chamber temperature. A linear guiding line through the origin shows the expected FWHM with the measured mixing chamber temperature and $\alpha_{PG}$. d) Time dependence of the resistive thermometer (black) and electron thermometer (blue) after switching off the mixing chamber heater at 150 mK. Both datasets are fit with an exponential decay.}}
    \label{fig:5}
\end{figure}

\begin{figure*}[!t]
    \centering
    \includegraphics[width=1.00\linewidth]{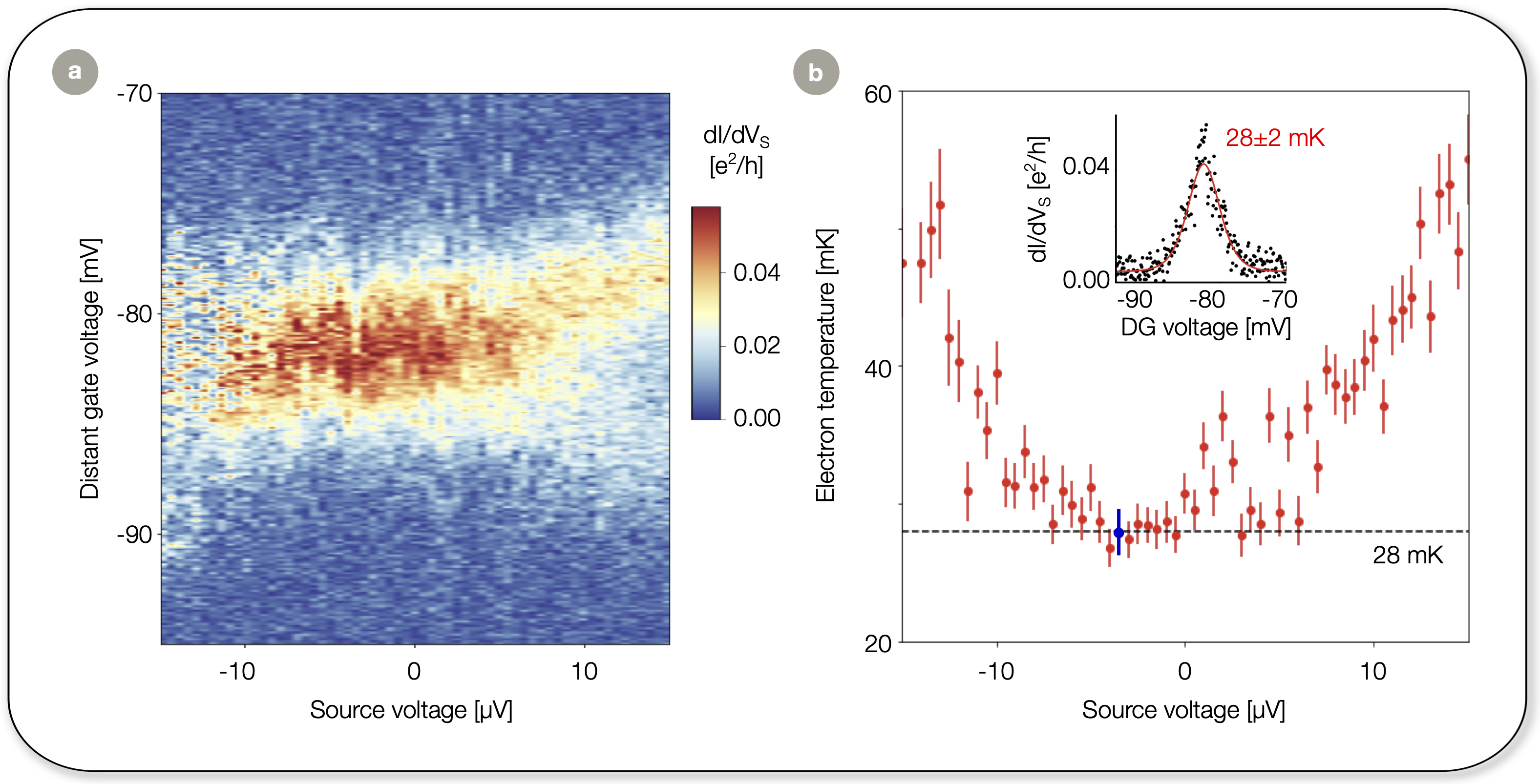}
    \caption{\textit{Electron thermometry measurement with further optimized parameters. a) High resolution scan of the Coulomb peak near zero bias, using a distant plunger gate and a voltage divider for the source voltage to increase the resolution. b) Measured electron temperature by fitting the Coulomb peaks with Eq. \ref{Eq3} (see inset for the fit resulting in the blue datapoint). A black dashed line shows the measured electron temperature of 28 mK.}}
    \label{fig:6}
\end{figure*}

\subsection*{Parameter extraction from quantum dot measurements}\label{sec:qd meas}
First, we extract the lever arm of the plunger gate $\alpha_{PG}$, which gives the relation between the energy level of the quantum dot and the voltage on this gate, given by the ratio of the plunger gate capacitance to the total capacitance, C$_{PG}$/C$_\Sigma$.
This ratio is directly given by the ratio between the height and width of the Coulomb diamond.
The height from zero bias to either the top or bottom of the diamond is given by the charging energy:
\begin{equation}\label{Eq2}
    E_C = \frac{e^{2}}{C_\Sigma} + \Delta E,
\end{equation}
where $e$ is the elementary charge and $\Delta$E is the level spacing of the quantum dot. Combined with the width of the Coulomb diamond given by E$_C$/$\alpha_{PG}$, the lever arm is found.
Secondly, these measurements can be used to estimate the relations between the different energy scales at play: the thermal energy of the electrons k$_B$T$_e$, the level spacing $\Delta$E, and the tunnel broadening h$\Gamma$.
Here k$_B$ is the Boltzmann constant, T$_e$ the electron temperature, h the Planck constant, and $\Gamma$ the total tunnel rate.
In the regime in which h$\Gamma$ $\ll$ k$_B$T$_e$ $\ll$ $\Delta$E, the best estimate of the electron temperature can be made from the single electron tunneling measurements.
For the condition $\Delta$E, k$_B$T$_e$ $\gg$ h$\Gamma$ to be true, the resulting conductance in the quantum dot must be much smaller than the conductance quantum 2e$^2$/h, which the differential conductance measurement can verify.
In the regime of h$\Gamma$ $\ll$ k$_B$T$_e$ $\ll$ $\Delta$E, the zero bias conductance peak is given by \cite{Beenakker91}:
\begin{equation}\label{Eq3}
    G = G_{max} \hspace{1mm} cosh^{-2}\left(\frac{\Delta\epsilon}{2k_BT_e}\right),
\end{equation}
where $\Delta\epsilon$ is the energy difference from the resonant tunneling condition, and G$_{max}$ is given by \cite{Beenakker91}:
\begin{equation}\label{Eq4}
    G_{max} = \frac{e^2}{4k_BT_e}\frac{\Gamma_{in}\Gamma_{out}}{\Gamma_{in}+\Gamma_{out}},
\end{equation}
With $\Gamma_{in}$ and $\Gamma_{out}$ the tunnel rates in and out of the quantum dot.
In contrast, $G_{max}$ does not depend on the temperature in the classical regime with k$_B$T$_e$ $\gg$ $\Delta$E.
Therefore, temperature dependent measurements of the conductance peak will verify, if indeed only a single quantum level influences the single electron tunneling (see Fig. \ref{fig:5}).

\subsection*{Data analysis}\label{sec:data}
From Fig. \ref{fig:4}c and \ref{fig:4}d, the plunger gate lever arm is measured as 0.074 ± 0.003.
The zero bias differential conductance peak at a plunger gate voltage of -424 mV in Fig. \ref{fig:4}d measures around 0.02 e$^2$/h, thereby verifying that we are in the regime $\Delta$E, k$_B$T$_e$ $\gg$ h$\Gamma$.
Furthermore, the amplitude of this peak depends on temperature, as shown in figure 5c, thereby also confirming k$_B$T$_e$ $\ll$ $\Delta$E. 
This figure shows that there is a good agreement between the temperatures measured by the resistive Ruthenium-oxide thermometer on the mixing chamber and the electron thermometry using the quantum dot at temperatures above 40 mK. 
Below 40 mK, the electron temperature is stabilizing, while the resistive thermometer still measures  a cooling of the mixing chamber plate.
Next, we show a measurement of the time constant of the electron thermalization.
Here we continuously measure the width of the zero-bias conductance peak as well as the Ruthenium-oxide thermometer for 4 hours after switching off the mixing chamber heater, as shown in Fig. \ref{fig:5}d.
Both thermalization curves are fitted with an exponential decay, for which we find a decay time $\tau_e$ of 9.8 minutes for the electron temperature measurement, which is close to the decay time $\tau_{MC}$ of 7.8 minutes for the resistive temperature sensor at the mixing chamber.

\subsection*{Further optimization}
Finally, improvements were made to the electron thermometry measurements, by finding the ideal values for several equipment settings. 
The ideal lockin amplifier amplitude was found to be 1 $\mu$V (created with an output of 100 mV and the use of a 1/10$^5$ voltage divider, as shown in Fig. \ref{fig:2}d), the ideal frequency to be 117 Hz, the ideal time constant to be 30 ms, and the ideal roll-off to be 12 dB.
Furthermore we use a 1/10$^3$ voltage divider for the source voltage (see Fig. \ref{fig:3}b), such that we can achieve a resolution below 10 nV using the QDevil QDAC.
To improve the resolution of the plunger gate, we use one of the distant gates (DG) further away from the dot instead, with a lever arm for the distant gate of 0.0016 ± 0.0001.
With these settings, we scan the source voltage close to the zero bias point, in order to make sure to find the narrowest point of the conductance peak (see Fig. \ref{fig:6}a).
We fit these conductance peaks around the zero-bias point with Eq. \ref{Eq3} (see inset Fig. \ref{fig:6}b) and find that the electron temperature stabilizes at 28±2 mK where the Coulomb peak is the narrowest (see Fig. \ref{fig:6}b).

\section{Conclusion}\label{sec:conclusion}
Using single electron transport in a GaAs/AlGaAs quantum dot, we performed electron thermometry measurements.
The experimental setup was specifically set up for lownoise measurements, of which a detailed description has been given. For the electron thermometry analysis, it has been established that the effect of tunnel broadening was negligible and that only a single quantum level influenced the differential conductance measurements.
After optimizing all experimental parameters, we find an electron temperature of 28 ± 2 mK, which we attribute to the use of a QDevil QFilter-II. 

\section*{Acknowledgments}\label{sec:acknowledgments}
The measurements shown in this application note were performed at the Center for Quantum Devices of the Niels Bohr Institute at the University of Copenhagen in Denmark.
A lot of appreciation goes to J. Kutchinsky, F. Kuemmeth, A. Kühle, M. von Soosten and S. Andresen for their invaluable input and the many useful discussions.
We like to thank F. Fedele for the fabrication of the here used GaAs quantum dot device, and F. Ansaloni, B. Brovang, and L. Lakic for setting up the cryogenic wiring.

\bibliography{bibliography/mybibfile}

\begin{figure*}[!h]
    \centering
    \includegraphics[width=1.00\linewidth]{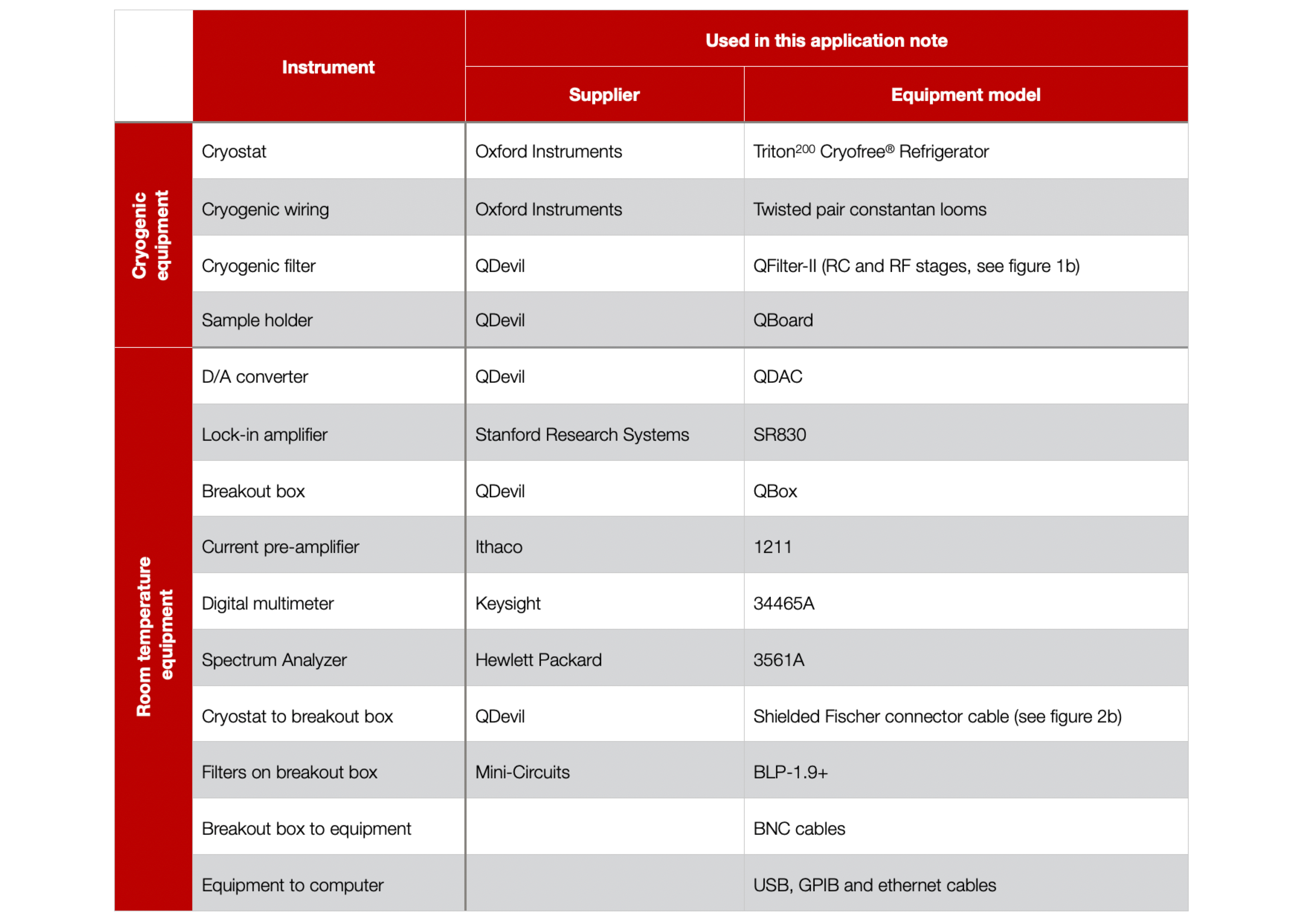}
    \caption*{\textbf{Appendix A.} List of equipment required for the Coulomb blockade electron thermometry measurement, noting the specific models used for the experiments in this application note.}
    \label{fig:A1}
\end{figure*}


\end{document}